\documentclass[]{aa} % for a printer version

\newcommand{\kms}{\,$\mathrm{km\, s^{-1}}$}

\newcommand{\msun}{$\mathrm{M_{\odot}}$}
\newcommand{\teff}{$T_{\rm eff}$}

\usepackage{color}
\usepackage{threeparttable}

\usepackage{txfonts}
\usepackage{graphicx}
\usepackage[authoryear]{natbib}
\usepackage{epsfig}
\usepackage{longtable}
\usepackage{lscape}

\usepackage{natbib}

\bibpunct{(}{)}{;}{a}{}{,} % to follow the A&A style

\begin{document}

\title{Lithium in M67: From the main sequence to the red giant branch}
 \subtitle{}

   \author{Giancarlo Pace\inst{1},       
           Matthieu Castro\inst{2},     
           Jorge Mel\'endez\inst{3},
            Sylvie Th\'eado \inst{4}
            \and Jos\'e-Dias do Nascimento Jr.\inst{2}
}

   \offprints{G. Pace, email: gpace@astro.up.pt}

\institute{Centro de Astrof\'isica, Universidade do Porto, Rua das Estrelas, 4150-762 Porto, Portugal
\and Departamento de F\'isica Te\'orica e Experimental, Universidade Federal do Rio Grande do Norte, CEP: 59072-970 Natal, RN, Brazil
\and Departamento de Astronomia do IAG/USP, Universidade de S\~ao Paulo, Rua do Mat\~ao 1226, S\~ao Paulo, 05508-900, SP, Brazil 
\and Laboratoire d'Astrophysique de Toulouse-Tarbes, Observatoire Midi-Pyr\'en\'ees, 31400 Toulouse, France
}

\date{Received : Accepted }

\abstract
% Context
{Lithium abundances  in open  clusters are a  very effective  probe of
  mixing  processes, and  their study  can help  us to  understand the
  large depletion of lithium that occurs  in the Sun. Owing to its age
  and metallicity,  the open cluster M67 is  especially interesting on
  this respect.  Many  studies of lithium abundances in  M67 have been
  performed,  but a homogeneous  global analysis  of lithium  in stars
  from subsolar masses and extending  to the most massive members, has
  yet  to be  accomplished for  a large  sample based  on high-quality
  spectra.}
% Aims: 
{We test our non-standard models,  which were calibrated using the Sun
  with observational data.}
% Methods:
{We  collect literature  data  to analyze,  for  the first  time in  a
  homogeneous   way,  the   non-local   thermal  equilibrium   lithium
  abundances of  all observed  single stars in  M67 more  massive than
  $\sim  0.9$ \msun.   Our  grid of  evolutionary  models is  computed
  assuming a non-standard mixing at metallicity ${\rm [Fe/H]} = 0.01$,
  using the Toulouse-Geneva evolution  code.  Our analysis starts from
  the entrance in the zero-age main-sequence.}
% Results:
{Lithium in  M67 is a  tight function of  mass for stars  more massive
  than  the Sun,  apart from  a few  outliers.  A  plateau  in lithium
  abundances is observed  for turn-off stars.  Both less  massive ($ M
  \leq 1.10$ \msun) and more massive  ($ M \geq 1.28$ \msun) stars are
  more depleted  than those  in the plateau.   There is  a significant
  scatter  in lithium  abundances  for  any given  mass  $M \leq  1.1$
  \msun.}
% Conclusions:
{ Our  models qualitatively reproduce  most of the  features described
  above,  although the  predicted  depletion of  lithium  is 0.45  dex
  smaller  than  observed  for  masses  in the  plateau  region,  i.e.
  between 1.1  and 1.28 solar masses.  More work is  clearly needed to
  accurately   reproduce  the   observations.    Despite  hints   that
  chromospheric  activity   and  rotation  play  a   role  in  lithium
  depletion,  no  firm conclusion  can  be  drawn  with the  presently
  available data.}

\keywords{Stars: fundamental parameters -- Stars: abundances -- Stars:
  evolution -- Stars: interiors -- Stars: solar-type}

\titlerunning{Lithium in M67: from the MS to the RGB}
\authorrunning{Pace et al.}

\maketitle

\section{Introduction}
\label{sec:Intro}

Lithium  is destroyed  by proton  capture at  temperatures  above $2.4
\times  10^6$  K.  In  Sun-like  stars,  material  in layers  at  this
temperature can  be mixed with photospheric  layers through convection
and diverse  processes such as diffusion,  meridional circulation, and
internal  gravity waves  \citep[e.g.][]{talon08}.   As a  consequence,
stars undergoing  these processes  have lower lithium  abundances than
their initial values.  This  phenomenon is known as lithium depletion,
and makes lithium abundance one of the most effective probes of mixing
processes in stars.

Astronomers  have long  dedicated huge  observational  and theoretical
efforts  to understand  how and  when  lithium is  depleted in  stars.
Depletion of lithium can occur during both the pre-main sequence (PMS)
and   the   main   sequence   (MS)  lifetime   of   solar-type   stars
\citep[e.g.][]{dm84,  dm94, dm03,  boe88, p89,  hobbs89, charbonnel92,
  pasquini94, cha95, ventura98, ms96, chen01, pt02, tv03b, lr04, tk05,
  Sestito05,  sestito06}.  Lithium  depletion  by rotationally  driven
mixing has been related to  the rotational history of stars, which, in
turn, has been  related to planet formation \citep[e.g.][]{bouvier08}.
The influence  of planets  is the subject  of much debate.   There are
claims of stronger lithium  depletion in planet-hosting stars but only
in  a  very  narrow   range  around  solar  \teff\  \citep{gonzalez08,
  israelian09}. However,  This evidence  has been challenged  by other
authors, who claim that no difference in lithium abundance is detected
between stars with and  without detected planets when unbiased samples
are analyzed \citep{melendez10, guezzi10, baumann10}.  Furthermore, at
a given mass and metallicity there is a clear relation between lithium
abundances and age,  independently of the star being  a planet host or
not \citep{melendez10, baumann10}.

Obviously, there  are two main parameters  favoring lithium depletion:
the depth, hence  the temperature to which the  mixing extends and the
length of  time for which the process  is active, i.e. the  age of the
star. However, the dependence of the former parameter on stellar mass,
age, and metallicity  is not simple, nor is  there a general consensus
on  which  additional  parameters  are involved.  Despite  decades  of
observations in clusters and field stars, no clear and widely accepted
picture has yet been drawn.

Lithium   abundance  in  open   clusters  is   a  precious   tool  for
understanding mixing  processes because  stars in open  clusters share
the same age  and initial chemical composition, therefore  allow us to
see  how  lithium depletion  varies  with  stellar  mass and  to  test
non-standard models  of stellar mixing.  A crucial  question, on which
observations of clusters can cast  light, is whether there is a spread
in lithium  abundance for stars with the  same fundamental parameters,
i.e.  mass,  chemical composition, and  age, which is to  say, whether
lithium abundance in  a given open cluster is a  tight function of the
stellar mass. In case it is  not, careful study of the deviating stars
would be important  to constrain the parameters that  cause the spread
of  lithium  abundances, such  as  for  example  the initial  rotation
velocity \citep{charbonneltalon05}.

Some early  observations indicated that  there is a certain  amount of
spread in  the lithium  abundances versus mass  distribution in  a few
clusters, and its absence, or at least the lack of evidence for it, in
many others.  Examples  of clusters belonging to the  former group are
the    Pleiades    \citep{duncan83,    soderblom93a}   and    Praesepe
\citep{soderblom93b}  for   low-mass  stars  ($M   \leq$  0.9  \msun).
However, many of the early claims about a spread in the lithium versus
mass  distribution   in  open  clusters  have   lately  been  revised.
\citet{xiong05, xiong06} found  that most, if not all,  of the lithium
dispersion for a given mass in members of $\alpha_{\mathrm{turb}}$ Per
and Pleiades, can be accounted for by inhomogeneous reddening, stellar
spots,  and  stellar  surface  activity,  rather  than  by  a  genuine
variation.  \citet{king00}, analyzing Praesepe stars, ascribed most of
the  scatter to  variations  in activity-regulated  ionization of  the
lithium  atom.  In  support  of this  hypothesis,  they mentioned  the
lesser amount  of scatter in  two older and  chromospherically quieter
open  clusters:  Hyades  \citep{thorburn93} and  M34  \citep{jones97}.
Now, we  can add to the  list several clusters older  than the Hyades,
such  as  NGC752   \citep{sestito04},  NGC188  \citep{randich03},  and
Berkeley32  \citep{randich09}.  There  are, however,  examples  of old
clusters presenting a  significant amount of scatter.  One  of them is
NGC3680, but the scatter in this cluster is especially at temperatures
higher  than $\sim$  6500 K  \citep{anthony09}, and  too few  stars at
lower temperature have been  analyzed.  Another old cluster apparently
showing a spread  in the lithium abundance distribution  is M67, which
is a  well-studied cluster, subject  to a number of  lithium abundance
investigations.   However,  a  comprehensive  study in  which  lithium
abundances  were  analyzed  for  a  large number  of  members  in  all
evolutionary  stages,  taking  advantage  of  already  available  high
signal-to-noise ratio (S/N) spectra, is still lacking.

Since M67 has an age  of about 3.9 Gyr \citep[e.g.][]{vdb07, castro11}
and  about solar  metallicity  \citep[e.g.][]{t00, randich06,  pace08,
  pasquini08, onehag11},  it can provide insight into  the behavior of
lithium  in relatively  old solar  type stars,  which is  important to
assess why some  old solar analogs in the field seem  to have too much
lithium for their ages \citep{baumann10}.

In  this work,  we homogenize  all previous  studies on  M67  stars by
obtaining a  consistent set  of temperatures, applying  corrections to
the  lithium   abundances  when  needed  (due   to  revised  effective
temperatures), and  taking into account  non-local thermal equilibrium
(non-LTE) effects on the lithium abundances. We analyze stars from the
MS to the red giant branch (RGB), comparing our models to the observed
behavior  of  lithium abundance  as  a  function  of mass.   In  Sect.
\ref{sec:Obs},  we describe  the data.   In Sect.   \ref{sec:Temp}, we
compare the  effective temperatures from different  references used in
the  paper.  In  Sect.  \ref{sec:Models},  we present  details  of our
stellar  models, and  in  Sect.  \ref{sec:Discussion}  we discuss  our
results.   Finally,   in  Sect.   \ref{sec:conclusion}   we  give  our
conclusions.

\section{Sample and data}
\label{sec:Obs}

Our  database  is  a  compilation  of literature  sources  of  lithium
abundance       measurements,      namely      \citet{cantomartins11},
\citet{castro11},        \citet{pasquini08},        \citet{randich07},
\citet{jones99}, and  \citet{bal95}.  When multiple  measurements were
available for a single object, we adopted the most recent. Some of the
data included  in our  compilation, namely data  from \citet{castro11}
and Table~2 in \citet{jones99},  are refinements of previous abundance
determinations.  The final sample amounts to 103 stars.  For the solar
twin  YBP~1194,  \citet{onehag11}  used  the highest  quality  spectra
available to  date to determine its lithium  abundance, although their
result is  essentially identical with that  of \citet{castro11}, which
is based on lower quality spectra;  thus, either of the sources can be
adopted to  measure the  lithium abundance of  YBP~1194.  The  data in
\citet{pasquini97},    \citet{garcialopez88},   \citet{spite87},   and
\cite{hobbs86} were  initially considered for  the present compilation
but eventually not used, since their whole sample was later reanalyzed
in some  of the  works cited above.   We also  did not include  in our
compilation  data  from  \citet{deliyannis94},  which is  a  study  of
tidally  locked binaries,  whose  mechanism of  lithium depletion  may
completely differ from that of single stars.

In some works, no quantitative estimations of the error in the lithium
abundances  were   given.   All  these  cases   are  discussed  below.
\citet{pasquini08}  stated that  all the  stars with  upper  limits in
their sample may have a lithium abundance comparable to the solar one.
Following  this piece  of information,  we raised  their  upper limits
(which in some cases were as low as 0.5 dex) to 1.0 dex, which is more
reasonable  considering  the  quality  of  the  spectra  available  to
\citet{pasquini08}.   However,  this  correction  hardly  changes  the
general picture  drawn using  data in Table~A.2  of \cite{pasquini08},
which points  to a significant  spread in lithium abundances  at about
one  solar  mass.  Their  estimation  of  the  uncertainty in  lithium
abundance due to  the uncertainty in the equivalent  width, for a star
with $A({\rm Li})=2.2$, is $\pm  0.04$ dex, and no evaluation is given
for stars with lower abundances.  Since lithium abundance is extremely
sensitive to temperature, the uncertainty in the latter may contribute
to  that of  the  former in  a  non-negligible way,  even  when it  is
extremely   low.    This    contribution   was   not   quantified   by
\citet{pasquini08}.  \citet{castro11}, in  reanalyzing the solar twins
of \citet{pasquini08} with spectral synthesis, found a global error of
0.1  dex.  Their  stars have  Li abundances  that are  lower  than the
average   of   the   sample   studied   by   \citet{pasquini08}.    We
conservatively  assumed  this  error   (0.1  dex)  for  all  stars  in
\citet{pasquini08}.

\citet{jones99} estimated  an error of  0.05 dex in  lithium abundance
due to the uncertainty in the equivalent width, and an equal error due
to the uncertainty  in the temperature of 50  K.  We were conservative
once  again, and we  summed linearly  these two  sources of  error and
assumed for  all lithium  abundances an uncertainty  of 0.1  dex. This
same  value   was  also  assumed   for  their  measurements   made  by
reanalyzing literature data.

Information on photometry and membership was taken, whenever possible,
i.e.   for 95  out of  103 stars,  from the  work  of \citet{yadav08},
otherwise  it was taken  from the  reference of  the study  on lithium
abundance.  Of the 95 stars  of our sample studied by \citet{yadav08},
87  had radial  velocity from  the same  source. For  the  remaining 8
stars, Yadav  et al.'s membership information was  based on photometry
and proper motions. Six of the 8 stars of the final sample not studied
by  \citet{yadav08},  have radial  velocity  data  available from  the
lithium abundance  study, therefore only  10 stars do not  have radial
velocities.   We discarded  stars with  radial velocities  outside the
range of  values from 30 to 38  \kms\ and removed from  our sample the
blue stragglers Sanders~2204 and Sanders~997 and other known binaries.

%%%%%%%% Figure 1 %%%%%%%%%%%%%
\begin{figure}[t!]
\begin{center}
\vspace{-0.1in}
\includegraphics[angle=0,height=9cm,width=9cm]{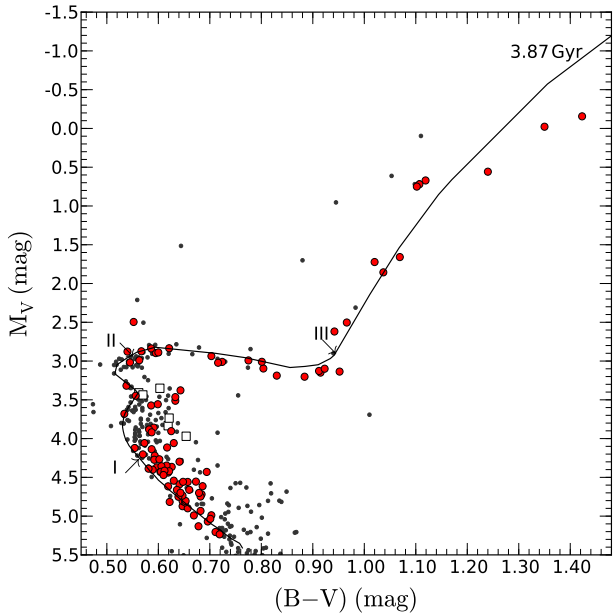}
\end{center}
\vspace{-0.2in}
\caption{Color-magnitude   diagram  for   M67   superposed  upon   the
  isochrone, which corresponds to an  age of 3.87 Gyr.  The red filled
  circles represent our work sample, i.e. stars with lithium abundance
  measurements.  Stars with  photometry available from \cite{yadav08},
  but with  no lithium abundances, are displayed,  for comparison with
  the isochrone,  as black points.  The squares  represent the deviant
  stars. The roman numbers I, II, and III are described in Sect. 5.1
  and correspond to the temperatures of, respectively, 6095, 6123, and
  4966 K.}
\label{fig:CMD}
\end{figure}
%%%%%%%%%%%%%%%%%%%%%%%%%%%%

In Fig. \ref{fig:CMD}, we plot a color-magnitude diagram of M67 stars,
and the isochrone calculated from  our models, with the parameters for
M67 found in \citet{castro11}, i.e. ${\rm  [Fe/H]} = 0.01$, $(m - M) =
9.68$, and $E(B - V) = 0.02$,  and age of 3.87 Gyr. Black dots are the
stars from the  sample of \citet{yadav08}, red filled  circles are the
stars studied in  this paper, and open squares  are deviant stars (see
Sect. \ref{subsec:deviant}).  Roman numbers I, II, and III are defined
in Sect.  \ref{sec:Discussion}.

Photometric  and  membership  information,  are available  at  CDS  in
electronic form (Table~1).

\addtocounter{table}{1}

\section{Temperatures and corrected lithium abundances}
\label{sec:Temp}

Lithium   abundances   are  strongly   sensitive   to  the   effective
temperature.   To check for  the dependence  on the  adopted effective
temperatures,  we compared different  estimates, using  the literature
source  from   which  the  lithium  abundances  were   also  taken,  a
photometric  calibration,   and  the  isochrone.    To  determine  the
effective temperature by  means of the isochrone, we  proceeded in the
following way:  we plotted in the  same CMD data points  for each star
and  the  isochrone computed  for  M67  age  and metallicity,  shifted
according  to its distance  modulus and  reddening.  For  each stellar
data  point, we  then took  the closest  point on  the  isochrone, and
adopted the corresponding physical parameters, i.e., in the case under
consideration, effective temperature. For  the remainder of the paper,
we refer to this kind  of temperature estimation as either ``isochrone
temperature'', or $T_{\rm iso}$  with the further specification of the
isochrone used.   ``Isochrone masses'' were computed in  the very same
way.  Summarizing, we have the following four temperature estimations:

\begin{enumerate}{}
\item the \teff\ adopted in the paper from which the lithium abundance
  was taken ($T_{\rm  orig}$);
\item the isochrone temperature adopting models and parameters for M67
  as   in  \citet{yadav08},   who   in  turn   adopted  BaSTI   models
  \citep{pietrinferni04} ($T_{\rm iso-BaSTI}$);
\item the isochrone temperature adopting isochrones and parameters for
  M67 in \citet{castro11} ($T_{\rm iso-Castro}$);
\item using  a photometric calibration,  from \citet{casagrande10} for
  dwarfs  and  sub-giants,  and  from \citet{kucinskas05}  for  giants
  ($T_{\rm calib}$  or ``calibration temperature'').
\end{enumerate}

A comparison between the  different temperature evaluations shows that
the differences  are significant.  When comparing  $T_{\rm orig}$ with
the  other temperature  evaluations,  the heterogeneity  of the  data,
especially in the assumed reddening for M67, is certainly a main cause
of these differences,  and one of the aims of the  present study is to
homogenize the different determinations.  The mean differences between
$T_{orig}$   on  the   one  hand,   and  $T_{\rm   iso-BaSTI},  T_{\rm
  iso-Castro}$, and  $T_{\rm calib}$ on the  other, are, respectively,
19, 50, and  95 K.  The differences between  calibration and isochrone
temperatures  arise, instead, from  the isochrone  in the  CMD leaving
most of  the points on the  right (cooler) side.  This  is because the
calibration  temperature is  affected  by the  presence of  undetected
companions, which move the data points brighter and cooler in the CMD.
In  computing isochrone  temperatures, we  exploit the  fact  that the
stars belong  to the same  cluster.  The isochrone  on the CMD  is the
locus where we  would expect all the photometric data  points to be if
there  were no photometric  errors and  no multiple  systems.  $T_{\rm
  iso}$ is  the temperature corresponding to this  locus, therefore it
should be  unaffected by the presence of  undetected companions.  This
is confirmed by  the overall good agreement between  $T_{\rm iso}$ and
$T_{\rm  orig}$ for  the  82 stars  studied in  \cite{cantomartins11},
\cite{castro11},  and  \cite{pasquini08}.   In  these  stars,  $T_{\rm
  orig}$  is based  on spectroscopic  analysis and  is  therefore very
precise.  We label $T_{\rm spec}$  this subset of $T_{\rm orig}$ based
on spectroscopic analysis.  The linear  best fit to the data points in
the $T_{\rm spec}$ versus isochrone temperature graph is very close to
the identity: $T_{\rm iso-BaSTI}=  0.99 \cdot T_{\rm spec}-2.5$ K.  We
can confidently  claim that the  errors in the  isochrone temperatures
are random in nature.

We note  that the two  different evaluations of  effective temperature
using  either   isochrone,  i.e.   $T_{\rm   iso-BaSTI}$  and  $T_{\rm
  iso-Castro}$, are  consistent with each other within  the margins of
error for most  of the stars.  The only  significant differences arise
for sub-giant stars and, to a lesser extent, for turn-off stars.

Corrections to  lithium abundance measurements  were applied according
to the  differences between $T_{\rm orig}$  and, respectively, $T_{\rm
  iso-BaSTI}$, $T_{\rm iso-Castro}$, and  $T_{\rm calib}$. At the same
time, we  also took into account  NLTE effects using the  grid of NLTE
corrections computed by \cite{lind09}.

%%%%%%%%%%%%%%%%%%%%%%%%%%%%
\begin{figure}[t!]
\begin{center}
\vspace{-0.1in}
\includegraphics[angle=0,height=9cm,width=9cm]{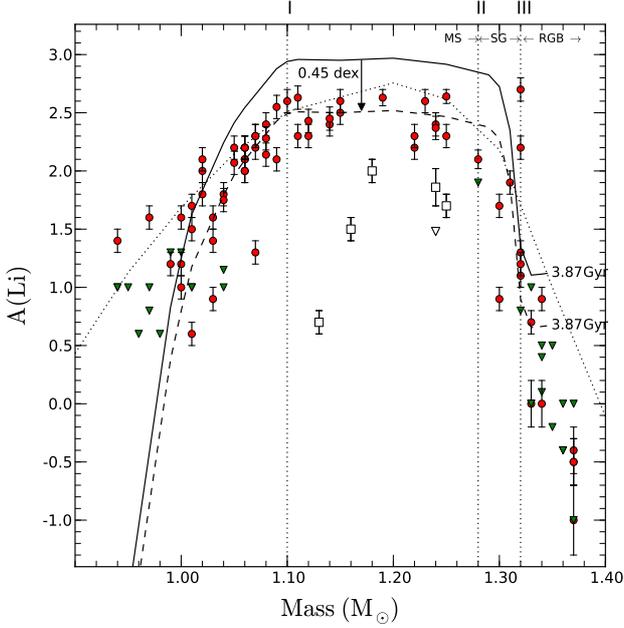}
\end{center}
\vspace{-0.2in}
\caption{Lithium abundances  as a function of  stellar mass determined
  from the isochrone. Red  filled circles represent stars with lithium
  abundance  redetermined. The green  triangles represent  the lithium
  abundance  upper limits.   The open  squares represents  the deviant
  stars as discussed in  Sect. 5.3. The continuous line corresponds to
  the lithium abundance predicted  by our models with rotation-induced
  mixing calibrated for the  Sun at  the age of  M67. The  dashed line
  represents the same model shifted  by 0.45 dex to match the observed
  lithium   plateau.    The  dotted   line   represents  models   with
  rotation-induced mixing calibrated on  the Hyades. The roman numbers
  I, II, and III are described in Sect. 5.1.}
\label{fig:liabmass}
\end{figure}
%%%%%%%%%%%%%%%%%%%%%%%%%%%%

All  temperature   estimations,  corrected  and   uncorrected  lithium
abundances, along with the photometric and membership information, are
available on-line in  Table~1. 

In Fig.   \ref{fig:liabmass}, we plotted  the lithium abundances  as a
function of mass for both the stellar data points and two models.  Red
filled  circles with  error  bars are  the  stars of  our sample  with
detected  lithium line,  and  green filled  triangles represent  upper
limits.  Open  squares with error  bars and, only  in the case  of one
upper limit, the open  triangle, represent the deviant stars discussed
in Sect.  \ref{subsec:deviant}.   The models are discussed extensively
in Sect.  \ref{sec:Models}.

As   mentioned  above,  \cite{cantomartins11},   \cite{castro11},  and
\cite{pasquini08}  computed  temperature by  carrying  out a  detailed
spectroscopic analysis.  Their  temperature evaluation is more precise
than that obtained using  photometric information.  Therefore, for the
82  stars studied  in these  works, we  used the  values given  in the
literature to  produce Fig.  \ref{fig:liabmass}. In  contrast, for the
remaining   21  stars,   we   used  $T_{\rm   iso-Castro}$,  and   the
corresponding corrected  value of the  lithium abundance.  For  all of
the stars, the mass adopted  is the isochrone mass, which was computed
using   the  isochrone  and   the  parameters   for  M67   adopted  in
\cite{castro11}.

\section{Stellar evolutionary models}
\label{sec:Models}

For this  study, stellar evolutionary  models were computed  using the 
Toulouse-Geneva  stellar   evolution  code  TGEC  \citep{huibonhoa08}. 
Details   on  the   physics  of   these   models  can   be  found   in 
\citet{richard96,richard04},        \citet{donascimento00},        and
\citet{huibonhoa08}.

\subsection{Input physics}

We used the OPAL2001 equation of state by \cite{rogers&nayfonov02} and
the  radiative opacities  by \cite{iglesias&rogers96},  completed with
the    low   temperature   atomic    and   molecular    opacities   by
\cite{alex&ferg94}.   The  nuclear  reactions  were described  by  the
analytical formulae of  the NACRE \citep{angulo99} compilation, taking
into account  the three \textit{pp}  chains and the CNO  tricycle with
the  \cite{bahcall&pin92} screening  routine.  Convection  was treated
according to  the \cite{bohmvitense58} formalism of  the mixing length
theory with  a mixing length  parameter $\alpha = l/H_{\rm  p} =1.74$,
where $l$  is the  mixing length and  $H_{\rm p}$ the  pressure height
scale.  For  the atmosphere, we  used a gray atmosphere  following the
Eddington relation,  which is a  good approximation for  MS solar-type
stars \citep{vandenberg08}.

The  abundance  variations  in  the following  chemical  species  were
computed individually in  the stellar evolution code: H,  He, C, N, O,
Ne, and Mg.  Both Li and Be were treated separately only as a fraction
of the initial abundance. The heavier elements were gathered in a mean
species Z. The initial composition follows the \cite{grevesse&noels93}
mixture with  an initial  helium abundance $Y_{\rm  ini} =  0.268$. We
chose to use the  ``old" abundances of \cite{grevesse&noels93} instead
of the ``new" mixture  of \cite{asplund09}.  This choice was motivated
by the  disagreement between models  computed with the  new abundances
and  the  helioseismic inversions  for  the  sound-speed profile,  the
surface   helium   abundance,    and   the   convective   zone   depth
\citep[e.g.][]{serenelli09}.       Furthermore,      according      to
\cite{caffau09},    the   solar    metallicity    using   their    own
three-dimensional  analysis  is given  by  the  values $Z=0.0156$  and
$Z/X=0.0213$,  which are closer  to those  of \cite{grevesse&noels93}.
In all cases, the solar abundances remain uncertain.  We note that the
accretion    of     metal-poor    material    \citep[e.g.][]{castro07,
  guzik&mussack10} as supported by  the lack of refractory elements in
the  solar  atmosphere  \citep{melendez09},  may help  to  reduce  the
discrepancy between the solar  model and helioseismic data.  Thus, the
low   solar   abundances   of   \cite{asplund09}   may   actually   be
representative of  the solar photosphere, while in  the solar interior
the abundances may be higher.
 
\noindent \textit{Diffusion and rotation-induced mixing.} \\

The   microscopic   diffusion  is   computed   with   the  atom   test
approximation.   All   models  include  gravitational   settling  with
diffusion  coefficients computed  as in  \cite{paquette86}.  Radiative
accelerations are not computed here, since we focus only on solar-type
stars where their effects are  small when mixing is taken into account
\citep{turcotte98, dela&pins05}.   Rotation-induced mixing is computed
as described  in \cite{tv03a}.  This  prescription is an  extension of
the approach of \cite{zahn92} and \cite{maeder&zahn98}, and introduces
the  feedback   effect  of   the  $\mu$-currents  in   the  meridional
circulation,   caused  by   the  diffusion-induced   molecular  weight
gradients.   It introduces  two free  parameters in  the computations:
$C_{\rm   h}$   and   $\alpha_{\rm   turb}$  \citep[cf.    Eq.    (20)
  of][]{tv03a}.   The evolution  in the  rotation profile  follows the
Skumanich's law  \citep{skumanich72} with an  initial surface rotation
velocity on  the zero-age main-sequence  (ZAMS) equal to $V_{\rm  i} =
100$  km.s$^{-1}$,  which roughly  corresponds  to  the mean  rotation
velocity of stars hotter than about  7000 K in the Hyades according to
the  statistical  study  of  \cite{gaige93}.  The  parameters  of  the
Skumanich's law are calibrated to match the solar rotation velocity at
the solar  age ($\sim  2 \ \mathrm{km.s^{-1}}$).   Other prescriptions
were  tested  by  other  authors  to model  the  lithium  destruction.
\cite{charbonneltalon99}   and   \cite{palacios03}  included   angular
momentum   transport   induced    by   mixing.    However,   since   a
rotation-induced  mixing alone  cannot account  for the  flat rotation
profile inside  the Sun, these  authors later introduced  the possible
effect  of internal  gravity  waves  triggered at  the  bottom of  the
convective  zone \citep[see e.g.][]{talon&charbonnel05},  which allows
the hot side of the  Li-dip to be reproduced.  Other authors suggested
that the internal magnetic field is more important than internal waves
in  transporting angular  momentum  \citep{gough&mcintyre98}.  In  any
case, when applied  to the solar case, all  of these prescriptions are
able to  reproduce the lithium  depletion observed in the  Hyades, and
the  results are  ultimately quite  similar \citep{talon&charbonnel98,
  tv03b}.

We  also include a  shear layer  below the  convective zone,  which is
treated as  a tachocline \citep[see][]{spiegel&zahn92}.  This layer is
parametrized  with an effective  diffusion coefficient  that decreases
exponentially downwards \citep{brun98, brun99, richard04}:
\begin{displaymath}
D_{\rm tacho} = D_{\rm bcz} \exp \left( \ln 2 \frac{r - r_{\rm bcz}}{\Delta} \right)
\end{displaymath}
where $D_{\rm bcz}$ and $r_{\rm bcz}$ are the value of $D_{\rm tacho}$
at the bottom of the convective  zone and the radius at this location,
respectively, and $\Delta$  is the half width of  the tachocline. Both
$D_{\rm bcz}$ and  $\Delta$ are free parameters and  the absolute size
of the tachocline (i.e., $\Delta/R_*$ where $R_*$ is the radius of the
star)  is   supposed  to  be   constant  during  the   evolution.   An
overshooting of  parameter $\alpha_{\rm  ov}=0.01 H_{\rm P}$  has been
included in the models that develop a convective core.

\subsection{Models and calibration}

We  calibrated a  model  of  1.00 \msun~to  match  the observed  solar
effective temperature and luminosity at the solar age. The calibration
method of  the models is  based on the  \cite{richard96} prescription:
for  a  1.00  \msun~star,  we  adjusted  the  mixing-length  parameter
$\alpha$ and  the initial helium abundance $Y_{\rm  ini}$ to reproduce
the  observed solar  luminosity  and  radius at  the  solar age.   The
observed  values that  we used  are those  of  \cite{richard04}, i.e.,
$L_{\odot}   =  3.8515  \pm   0.0055  \times   10^{33}$  erg.s$^{-1}$,
$R_{\odot} = 6.95749 \pm 0.00241  \times 10^{10}$ cm, and $t_{\odot} =
4.57 \pm  0.02$ Gyr.  For the  best-fit solar model, we  obtained $L =
3.8501 \times  10^{33}$ erg.s$^{-1}$ and $R =  6.95524 \times 10^{10}$
cm at an age $t = 4.576$ Gyr.

The  free  parameters of  the  rotation-induced  mixing determine  the
efficiency of the  turbulent motions.  They are adjusted  to produce a
mixing  that is  both:  1) efficient  and  deep enough  to smooth  the
diffusion-induced helium  gradient below the  surface convective zone,
thus improving the agreement between the model and seismic sound speed
profiles;  2) weak  and shallow  enough  to avoid  the destruction  of
Be. Following \cite{gs98}, the Be abundance of the Sun is $A({\rm Be})
= 1.40 \pm  0.09$. We obtained a slight Be destruction  by a factor of
1.33 with  respect to the meteoritic  value, which is  well within the
error in the solar Be abundance.

The calibration of the tachocline allows us to reach the solar lithium
depletion ($A({\rm Li}) =  1.10 \pm 0.10$ \citep[e.g.][]{gs98} and for
our best-fit  solar model we obtained  $A({\rm Li}) =  1.04$.  We also
checked  that the  sound velocity  profile  of our  best-fit model  is
consistent  with  that  deduced  from  helioseismology  inversions  by
\cite{basu97}.   Our  calibration   is  in  excellent  agreement  with
helioseismology, more accurately than 1\% for most of the star, except
in the deep interior, where the discrepancy reaches 1.5\%.

We computed  a grid of  evolutionary models of  masses in the  0.90 to
1.34  \msun\  range,  with a  step  of  0.01  \msun\  in mass,  and  a
metallicity of  ${\rm [Fe/H]}  = 0.01$, which  is a simple  average of
different     estimates     of      the     metallicity     of     M67
\citep[see][]{pasquini08}. We ran the models  from the ZAMS to the top
of the  RGB for the most  massive stars. The input  parameters for all
the models are the same as for the 1.00 \msun \ model.

There is no  a priori reason why the calibration  of the parameters of
the extra  mixing for the model of  1.00 \msun \ should  also hold for
different masses, and  at different times during the  evolution of the
stars. In  their work, \citet{tv03b}  adjusted the free  parameters of
the rotation-induced mixing  with the TGEC for each  one of the models
of different masses to obtain the correct lithium depletion at the age
of the Hyades. They noted that the horizontal diffusion coefficient is
very mass-dependent.  The uncertainty associated with these parameters
led us  to analyze  their impact  on the destruction  of lithium  as a
function  of stellar  mass.   To do  so,  a first  set  of models  was
calculated to reproduce the profile of lithium abundance as a function
of  mass  for  the  Hyades  open  cluster.  Our  Hyades  sample  is  a
compilation of  EW measurements  of the lithium  line at 6708  AA from
several  sources  \citep{randich07,  thorburn93,  soderblom90,  boe88,
  BT86, duncan83, rebolo88}.  All  the stars identified as binaries in
\cite{thorburn93}  were excluded.   When more  than one  reference was
available  for  a   star,  we  used  the  most   recent.   Masses  and
temperatures were taken either directly from \cite{bal95}, or computed
from  the  relationship  between   B-V  and,  respectively,  mass  and
temperature, obtained from  the data in the same  paper. The result of
the  compilation is  available at  CDS in  electronic  form (Table~2).
\addtocounter{table}{1}    The   parameters   $C_{\mathrm    h}$   and
$\alpha_{\mathrm{turb}}$ of the rotation-induced mixing of the models,
with masses from 0.800 to 1.400 \msun, with a step of 0.050 \msun, are
calibrated  to  reproduce at  the  age of  the  Hyades  (625 Myr)  the
observed destruction  of lithium.  In  these calibrations, we  did not
include the  tachocline that  is used in  the code to  calibrate solar
models, because in  the solar mass stars the  bottom of the convective
zone  is very close  to the  lithium destruction  layer \citep{tv03b}.
The same calibration is then used  to calculate a set of models of M67
stars.    The   values  of   the   parameters   $C_{\mathrm  h}$   and
$\alpha_{\mathrm{turb}}$  for  all  calibrations  are given  in  Table
\ref{tab:calib}.

\begin{table*}
\caption{Parameters of the rotation-induced mixing in TGEC models.}
\label{tab:calib}
\centering
\begin{minipage}[t]{\textwidth}
\centering
\begin{tabular}{c|c|cc}
\hline
Calibration & Mass (\msun) & $C_{\mathrm h}$ & $\alpha_{\mathrm{turb}}$ \\
\hline
Sun  & 1.000 & 9000 & 1.00 \\
\hline
 & 0.800 & 3400 & 4.00 \\
 & 0.850 & 10000 & 5.00 \\
 & 0.900 & 10000 & 5.50 \\
 & 0.950 & 10000 & 5.60 \\
 & 1.000 & 10000 & 5.15 \\ 
 & 1.050 & 7000 & 4.25 \\
 Hyades & 1.100 & 5000 & 3.30 \\
 & 1.150 & 4500 & 2.50 \\
 & 1.200 & 2500 & 2.10 \\
 & 1.250 & 2546 & 2.15 \\
 & 1.300 & 2600 & 2.75 \\
 & 1.350 & 6500 & 3.50 \\
 & 1.400 & 10000 & 4.50 \\
\hline
\end{tabular}
\end{minipage}
\end{table*}
 
\section{Discussion}
\label{sec:Discussion}

The observed spread in  the lithium abundance versus mass distribution
of  M67 for  stars of  about one  solar mass  or lower,  which  can be
clearly  seen  in  Fig.  \ref{fig:liabmass}, suggests  that  an  extra
variable  (in addition  to mass,  age,  and metallicity)  is at  work,
leading to  different lithium depletion  in cluster stars of  the same
mass.  The fundamental question here  (since the spread is present) is
which other variable may cause a  large range of lithium abundance at a
fixed mass?

Our database includes  stars ranging from the MS  to the sub-giant and
giant branches, so is well-suited  to both address the issue of spread
and test models of non-standard mixing for different masses.

Models that point  to rotational mixing as the  major cause of lithium
depletion  \citep[see e.g.][  for  a review]{pinsonneault10},  predict
that this only depends on  the age, metallicity, mass, and the initial
angular momentum of  the star. This last parameter,  however, may vary
for  otherwise similar  stars,  introducing a  spread  in the  lithium
abundances.   The variation  in  the initial  rotation  velocity by  a
factor  two in  our models  has a  weak effect  of about  0.05  dex of
magnitude  for the  lithium destruction,  because the  Skumanich's law
used implies that there is a strong drop in the rotation velocity very
early on in stellar evolution.   The Skumanich's law is empirical, and
we should in  the future include in the code  the transport of angular
momentum to  ensure a  more reliable estimation  of the spread  of the
lithium  abundance due  to a  possible spread  in the  initial angular
momentum.

The  depth of  the surface  convection  zone and  the nuclear  burning
depend strongly on the total stellar mass \citep{donascimento09}, thus
the most  appropriate way to  investigate the abundance of  lithium in
these  M67 stars  is first  to  divide our  sample of  stars in  three
different  groups of mass  range. MS  stars with  $M \leq$  1.1 \msun,
stars close to  the turnoff with 1.1 \msun $< M  \leq$ 1.28 \msun, and
evolved sub-giant and giant stars with $M >$1.28 \msun.

\subsection{Lithium abundance and mass}
\label{sec:liandmass}

In Fig.  \ref{fig:liabmass}, we compare our model predictions with our
inferred stellar masses  and lithium abundances for our  sample of M67
stars.  The  solid line represents  an isochrone constructed  with all
the models  including the  effects of atomic  diffusion, gravitational
settling,  and  rotation-induced  mixing  calibrated for  the  Sun  as
described in Sect.~\ref{sec:Models}. The age of the models is 3.87 Gyr
as  determined in \citet{castro11}.   Initial lithium  abundances were
chosen  to equal  $A({\rm Li})  = 3.26$,  the estimated  initial solar
value based  on meteorites \citep{asplund09}.   This lithium abundance
is  0.05 dex  lower than  the previously  accepted  meteoritic lithium
abundance \citep[$A({\rm Li}) =  3.31$, see][]{gs98}.  The dashed line
represents  an isochrone  for the  same models  but where  the initial
lithium abundance  was rescaled by $-$0.45  dex to fit  the plateau at
$A({\rm  Li}) =  2.5$.  This  suggests  that the  initial M67  lithium
abundance may  have been lower than  3.26, or that our  models are not
depleting  enough  lithium.   We  note,  however,  that  according  to
\citet{deliyannis94}, based on a short-period tidally locked binary in
M67, the initial lithium abundance in  M67 was at least $A({\rm Li}) =
3.0$. This would  reduce the shift necessary to  match our models with
the lithium content in the plateau to only about $-$0.2 dex.  However,
the  hypothesis of  a lower  initial  lithium content  seems weak  and
unrealistic. We  are left with an insufficient  destruction of lithium
by rotationally  driven mixing mechanisms.   \citet{tv03b} showed that
for  masses higher than  1.0 \msun,  it is  necessary to  increase the
values of  these parameters as a  function of mass to  account for the
lithium depletion at the age  of the Hyades.  To analyze the influence
of these parameters on  the lithium destruction, we calculated another
set  of models, in  which the  rotation-induced mixing  parameters was
calibrated for each mass,  as described in Sect.~\ref{sec:Models}. The
isochrone at  the age  of M67  of these models  is represented  by the
dotted line  in Fig.~\ref{fig:liabmass}.  These  models reproduce more
closely in  a quantitative way  the profile of lithium  destruction in
M67, but the  arbitrary calibration of the mixing  parameters for each
mass is  a very  unsatisfactory method. There  appears to be  no clear
relation  between the  mixing parameters  and mass  that might  have a
physical significance.

\ {\bf For stars with masses  lower than 1.1 \msun} (roman number I in
Fig. \ref{fig:CMD} and Fig. \ref{fig:liabmass}), the lithium depletion
progressively increases toward lower masses.  A dispersion is observed
in this mass range and confirmed by several authors \citep{pasquini97,
  jones99,  randich02, randich07,  pasquini08}. We  note that  in this
region, some  upper limits used  may be considered as  too optimistic.
In this context, \cite{onehag11}  analyzed the M67 solar twin YBP~1194
using spectra of  good quality ($R \sim 50,000, S/N  = 160$), but even
with such  a spectrum  they found that  it is challenging  to robustly
determine a lithium  abundance in the solar twin  YBP~1194.  A careful
analysis  of  solar analogs  in  M67  using  high quality  spectra  is
urgently needed for stars around  1 \msun.  However, we note that four
stars at 1.01 \msun\ span as wide a range of lithium abundances as 1.1
dex, and one of them is  a revised upper limit.  A considerable amount
of scatter in the data in  this region remains after raising the upper
limits, whose real presence should be considered highly probable based
on the presently available data.   Stellar rotation and the history of
angular momentum  induce it either directly  through rotational mixing
or  indirectly  by  driving  other  processes such  as  diffusion  and
internal gravity waves \citep{ms00}.   Planetary accretion may also be
at  work  in solar-type  stars,  and  induce mass-independent  lithium
destruction \citep{theado10}.
 
The  low lithium  abundance observed  in  low-mass stars  in M67  also
exists because  these stars have had  more time to  burn their lithium
during the PMS  and have a surface convection  zone that both retreats
more slowly and ends up at greater depths on the MS. We note that, for
this  range of  masses  and these  evolutionary  stages, the  standard
models  predict that  the bottom  of the  convective zone  is  not hot
enough  to  account for  the  decrease  in  lithium, even  though  the
observed low lithium abundances clearly  indicates the need for a more
realistic representation of transport mechanisms in low mass stars.

\ {\bf  Stars with masses  1.1 \msun $<  M \leq$ 1.28  \msun} (between
roman number I and  II in Figs.  \ref{fig:CMD} and \ref{fig:liabmass})
are close  to the turn  off (or  about to leave  the MS) and  at these
masses  the surface  convection zone  and the  mixed layers  below the
convection zone  are too thin  to support temperatures high  enough to
cause the  nuclear destruction of  lithium.  Furthermore in  this mass
range the  PMS lithium depletion  tapers off to essentially  zero.  We
can therefore easily  see why there should be a  plateau with a nearly
constant lithium abundance for this mass range.  One important test of
our model will  be in fact to reproduce the  general morphology of the
lithium depletion behavior.

This  is  the   region  where  the  offset  between   our  models  and
observations becomes apparent, as discussed above.

\ {\bf Stars with masses above 1.28 \msun} are sub-giants following an
evolutionary   path    from   the    turn-off   to   the    RGB.    In
Figs. \ref{fig:CMD}  and \ref{fig:liabmass},  this stage is  marked by
the roman  numbers II  and III.  The  lithium abundance  predicted for
standard models of  sub-giant and giant stars is  mainly controlled by
dilution \citep{iben66, iben67a,  iben67b, scalo80}.  This process was
described originally by \citet{iben65} and appears when a star evolves
off the  MS.  When the stars  crosses the sub-giant  branch and climbs
the RGB, the  surface convective zone increases its  fraction in mass.
Lithium-poor material rises  to the surface and is  mixed with Li-rich
material.   This process stops  when the  convective zone  reaches its
maximum   size.    From   our   models,  the   predicted   value   for
lithium-abundance  post-dilution  is $A({\rm  Li})  =  1.0$ dex.   The
observations  clearly  show  the  need for  a  non-standard  transport
processes    for    sub-giants    and    evolved   stars    of    M67.
\citet{cantomartins11} discussed the  depletion of the surface lithium
abundance and  whether there  is an additional  non-standard transport
process related  to the  transport of matter  and angular  momentum by
meridional circulation and shear-induced turbulence \citep{ryan95}.

\subsection{Prediction by other models }

\begin{figure}[t!]
\begin{center}
\vspace{-0.1in}
\includegraphics[angle=0,width=9cm]{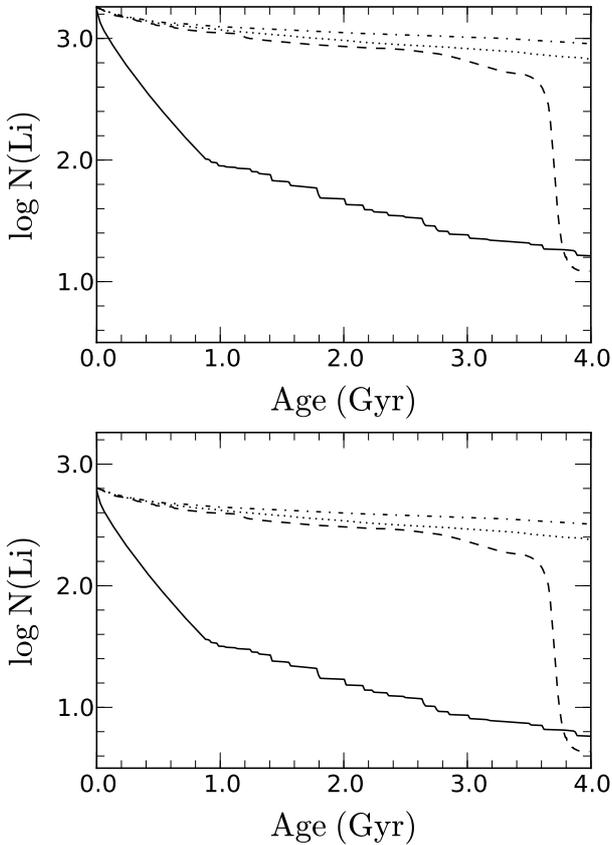}
\end{center}
\vspace{-0.2in}
\caption{Lithium abundance  predicted by our  models as a  function of
  time for different masses.  {\it Upper  panel}: for a MS star with M
  = 1  \msun (solid line),  stars of M  = 1.15 \msun  (dot-dashed) and
  1.28  \msun  (dotted) currently  around  the  turn-off,  and a  more
  massive  star  of  1.33  \msun  (dashed  line)  that  is  already  a
  giant. {\it Lower panel}: same plot, but after a -0.45 dex shift was
  applied (as in Fig. \ref{fig:liabmass}). }
\label{fig:liage}
\end{figure}

Our models successfully  reproduce qualitatively the lithium depletion
seen  in   different  mass   regimes  in  M67   stars,  as   shown  in
Fig. \ref{fig:liabmass}.  In Fig.  \ref{fig:liage}, the evolution with
age  at different  masses is  shown.  We  note that  fully independent
non-standard models  by \cite{xd09}, also show  similar trends between
lithium depletion, mass, and age.

As  shown in  Fig.   1 of  \cite{xd09}, at  4  Gyr, there  is a  large
depletion in  the lithium for models  of 1 \msun. As  in our models,
the less depleted stars are those with masses somewhat higher than 1.1
\msun, although  the \cite{xd09}  model also fails  to predict  exactly the
peak  or plateau  in  lithium abundance  for  M67 stars  close to  the
turnoff,  requiring probably  an additional  lithium depletion  of 0.3
dex. Stars with masses higher than 1.3 \msun\ show severe depletion at
4  Gyr, probably  because they  should be  in the  sub-giant  or giant
phase. Thus,  qualitatively, the models by  \cite{xd09} also reproduce
the M67 observations.

\subsection{Deviant stars}
\label{subsec:deviant}

Although overall lithium  abundance appears to be a  tight function of
mass  for stars  more  massive than  1.06  \msun,  a few  objects
dramatically depart from the general trend.

These objects are marked by open  square symbols in Figs. \ref{fig:CMD} 
and \ref{fig:liabmass}, and are discussed below.

{\bf YBP 1075,  YBP 750, YBP 769}: These  objects, despite being quite
distant  from the isochrone  in the  CMD diagram  (more than  0.1 mag,
limit under which 80\% of the sample lies), are all secure members (99
or 100\%  of probability,  based also on  radial velocity).   They are
therefore  likely to  have companions  too small  to be  detected, but
bright enough to affect color and brightness.  Their lithium abundance
deficiency, with  respect to  the mean trend  with the mass,  is about
2.3,  1.0,  and 0.5,  dex  respectively,  much  more than  errors  can
possibly explain, and  they have estimated masses in  the narrow range
between 1.12  and 1.18  \msun. Therefore, the  question of  whether we
identify a  ``secondary dip''  is legitimate, although  strong caution
should be exercised before claiming it as highly probable.  Their mass
range encompasses  four more stars,  three of which  lie significantly
closer to the isochrone.  If we formulate the intriguing hypothesis of
an extra-mixing mechanism triggered in this narrow mass range by a low
mass  companion,  we are  still  left  with  the unexplained  case  of
YBP~1680, which is  also likely to have a  small undetected companion,
but  does  not  show  any  anomaly  as far  as  lithium  abundance  is
concerned.  This matter deserves further attention.

{\bf YBP 890}: This star is also a secure member, but, contrary to the
three discussed  above, its  distance to the  isochrone in the  CMD is
only  0.03 mag.  It  has a  mass of  about 1.24  \msun, and  a lithium
abundance about 0.9  dex lower than YBP~961, which  is estimated to be
at about  the same mass. Its  lithium abundance is an  upper limit, it
may  therefore  be more  peculiar  than  it  appears.  This  star  has
undoubtly undergone a strong amount of extra mixing.

{\bf  YBP~871 and  YBP~778} are  also significantly  under-abundant in
lithium, and have masses similar to that of YBP~890. Their distance to
the isochrone in  the CMD is about 0.06 mag, which  is larger than for
YBP~890 but not dramatically large.   In their case, however, we still
cannot completely rule out the  possibility of a combination of errors
in  the abundance  and  mass  to explain  their  peculiar position  in
Fig. \ref{fig:liabmass}.

{\bf YBP 942}:  This is a secure member of the  cluster, the only star
with a  peculiarly high  lithium abundance, which  is 0.5 and  0.8 dex
higher   than   the  slightly   less   massive   stars  YBP~1320   and
YBP~1318. Rather than an error  in the abundance, an overestimation of
the mass  of this star of about  0.05 \msun\ could help  us to explain
its abundance anomaly. However, its distance from the isochrone in the
CMD is large at about 0.1 mag.

%%%%%%%%%%%%%%%%%%%%%%%%%%%%
%%%%%%%%%%%%%%%%%%%%%%%%%%%%

\subsection{Lithium abundance, rotation, and chromospheric activity in M67}

In   stars  with   a   convective  envelope,   the  chromosphere   and
chromospheric   activity  are   generated  by   non-radiative  heating
mechanisms  caused  by  a  magnetic  field  \citep[see  e.g.][  for  a
  review]{hall08}.  Rotation and differential rotation play a key role
in this  process, because they  are at the  base of the  dynamo effect
that   sustain  magnetic  activity   \citep{bohmvitense07}.   Magnetic
braking  spins  down the  star,  which  therefore loses  chromospheric
activity as  it loses angular momentum.   While until a  few years ago
there  was  general   consensus  that  chromospheric  activity  decays
smoothly       during      the      whole       stellar      life-time
\citep[e.g.][]{skumanich72,     barry87,    soderblom91,    donahue98,
  lachaume99},  which is  still  widely accepted  \citep[]{mamajek08},
alternative views  have also been expressed  later on \citep[]{pace09,
  lyra05,  zhao11}.  However,  nobody questions  that young  stars are
more  chromospherically   active  than   stars  older  than   1.5  Gyr
\citep{soderblom10}.

This picture  suggests that chromospheric  activity may be  related to
rotation and age, in a very  similar way as lithium abundance. This is
the rationale for a comparison between these three parameters in M67.

\begin{figure*}
\begin{center}
\begin{tabular}{c c}
\includegraphics[angle=0,width=8cm]{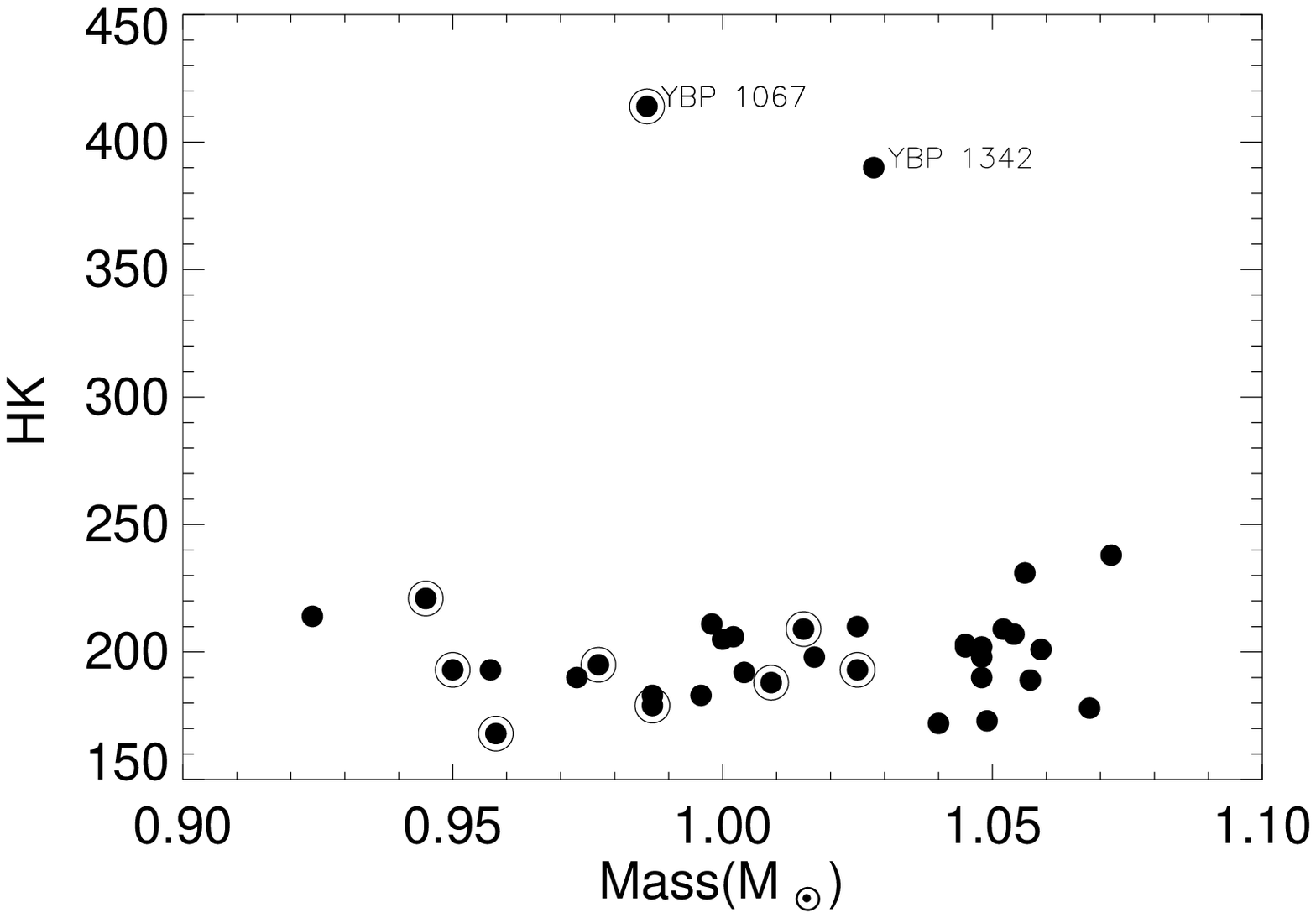}&
\includegraphics[angle=0,width=8cm]{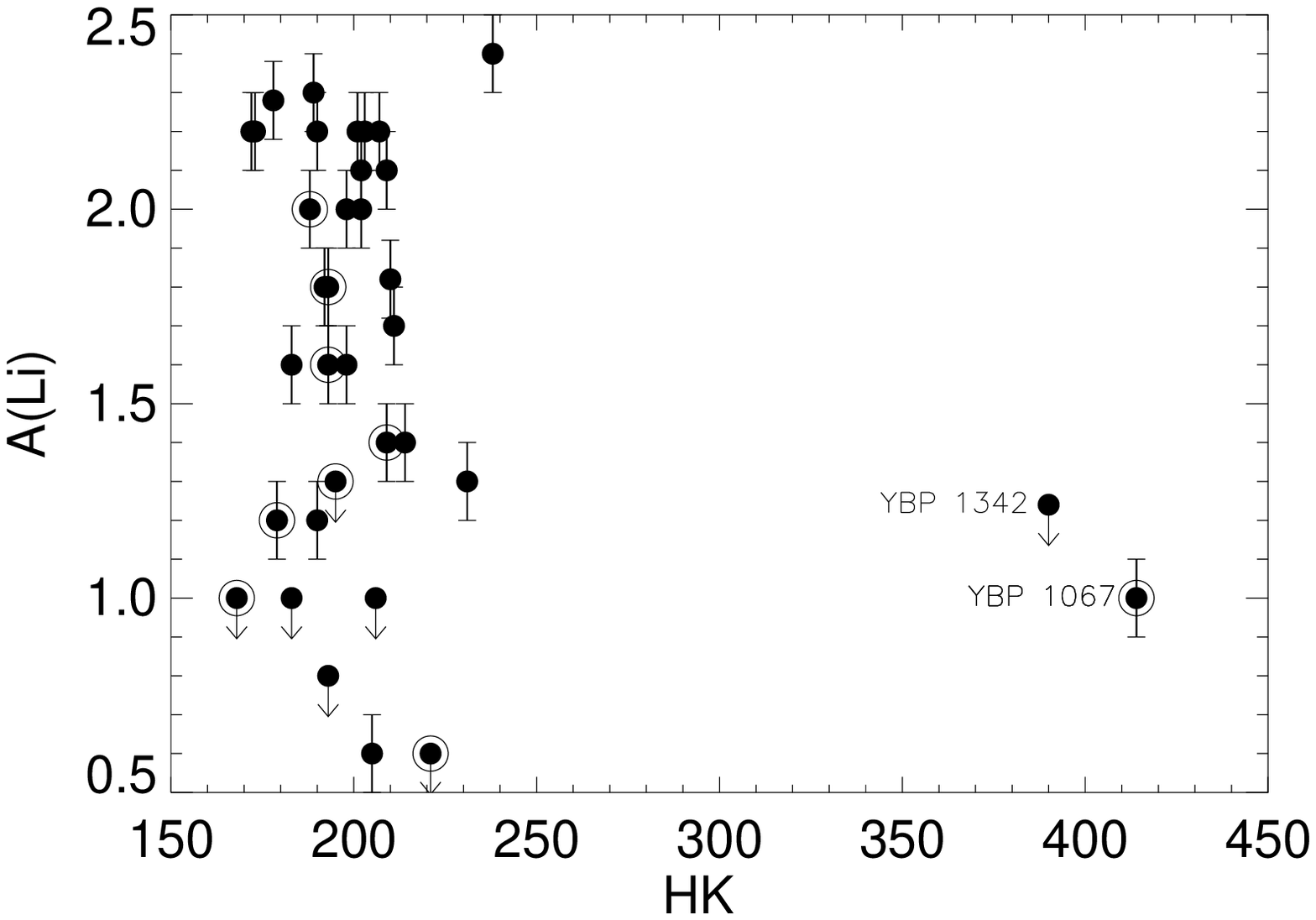}\\
\end{tabular}
\vspace{-0.2in}
\caption{On the left: chromospheric activity as a function of mass. On
  the right:  lithium abundances versus  chromospheric activity. Empty
  circles  around  filled symbols  indicate  the  stars with  measured
  projected rotation  velocity. Among these measurements, only that of
  YBP~1067 exceeds 2 \kms, and it is of 4 \kms.}
\label{fig:cromact}
\end{center}
\end{figure*}

The most  extensive chromospheric-activity survey made in  M67 is that
by \cite{giampapa06}, who  collected data for over 60  MS stars.  They
used  the  HK   index,  i.e.   a  measure  of   the  strength  of  the
chromospheric emission in the core of the Ca {\rm II} H and K line, in
m{\AA}.  The HK index is  neither transformed into flux, nor corrected
for  photospheric contribution.   \cite{mamajek08} transformed  the HK
indices  of \citeauthor{giampapa06}  into $R^{\prime}_{\rm  HK}$, i.e.
they  subtracted the  photospheric contribution,  transformed  it into
flux, and  normalized it for  the bolometric emission.  The  data were
taken over a  time span of five years, and  measurements for each star
were  time-averaged over  large part  of  this interval,  which is  of
crucial  importance  since  the  chromospheric  activity  undergoes  a
short-term variation  analog of the 11-year  solar cycle.  Thirty-five
stars  studied  by   \citeauthor{giampapa06}  have  published  lithium
abundances   that   are   collected   here   and   plotted   in   Fig.
\ref{fig:cromact}. In this figure, we show the HK index in m{\AA} as a
function  of mass  ({\it  Left  panel}) and  HK  index versus  lithium
abundances ({\it  Right panel}).  Here  upper limits are  indicated as
downward arrows.  Circles surrounding  the dots indicate that the star
has a  rotation velocity measurement available (see  below).  The left
panel has the  purpose of identifying the two  active outliers.  It is
similar to Fig.  4 in \citeauthor{giampapa06}, in which B-V color is a
substitute for  mass, but we note  that among the two  active stars in
both figures,  the only star in common  is Sanders~1050/YBP~1342.  The
star Sanders~747/YBP~681 is given  in Fig.  4 of \citeauthor{giampapa06}
and not in  our Fig. \ref{fig:cromact} because it is  a binary, and we
excluded  it   from  our  compilation.   On  the   contrary,  we  plot
Sanders~1452/YBP~1067,  which is  outside the  margins of  Fig.   4 of
\citeauthor{giampapa06}.  In  the {\it right panel}, we  show that the
two active outliers are  highly lithium-depleted.  However, these high
levels  of lithium  depletion are  also found  in many  other inactive
stars.  The  Pearson correlation coefficient between the  HK index and
the lithium abundance  is weak but significant at  $-$0.25.  This weak
correlation cannot be an effect  of a correlation between HK index and
mass, which,  for our sample, are completely  unrelated.  However, the
correlation between HK index and  lithium abundance is entirely due to
the two  chromospherically active  and highly lithium  depleted stars.
If we remove them, the correlation between HK and lithium abundance is
$-$0.05,  which corresponds  to a  probability of  the  two quantities
being unrelated of about 80\%.

\cite{reiners09}  measured  the projected  rotation  velocities of  15
stars selected from the  sample of \citeauthor{giampapa06} in order to
check whether  rotation was  at the base  of the  higher chromospheric
activity  level of  some stars.   The measurements  were based  on the
cross-correlation  profile of high-resolution  spectra. Nine  of these
stars are  present in our  compilation of lithium abundances,  and are
indicated  in Fig.   \ref{fig:cromact} with  a circle  surrounding the
dot.   With the  only  exception  of YBP~1067,  all  of the  projected
rotation velocities  are consistent with  the solar rotation,  as they
have an upper  limit of about 2 \kms. On the  contrary, YBP~1067 has a
projected rotation velocity of about  4 \kms, i.e. it rotates at least
twice  as fast  as  the  sun.  Whether  this  suggests a  relationship
between rotation  and lithium abundance  is unclear. On the  one hand,
YBP~1067 has a very low lithium abundance, on the other hand there are
two slow rotators with an even lower lithium content and one more slow
rotator with an upper limit  only slightly above the lithium abundance
measurement for  YBP~1067.  More data  are warranted to shed  light on
this matter.  In particular, it would be very important to obtain true
rotation periods instead of only $v$ sin $i$.

\section{Conclusions}
\label{sec:conclusion}

Spectroscopic observations  of stars in  the open cluster  M67 provide
important   constraints  to  test   non-standard  models   of  lithium
depletion.  We have rederived  the effective temperatures of M67 stars
and corrected  the lithium abundances available in  the literature, in
order  to  have  a   homogeneous  data-set  for  comparison  with  our
models.  We have  also taken  into account  NLTE effects,  providing a
homogeneous set of NLTE Li  abundances in M67.  Tables with our \teff,
NLTE lithium abundances, and masses are presented, so that other groups
can test their non-standard stellar models.

M67 stars close  to the turn-off, with masses $M =  1.1 - 1.28$ \msun,
have a peak or a plateau in lithium abundances. Less massive stars ($M
< 1.1$  \msun) display strong  lithium depletion, which  increases for
lower masses  owing to a  deepening of their convection  zones.  Evolved
sub-giant  and  giant stars  with  $M >  1.28$  \msun\  show lower  Li
abundances due to  a dilution of their original  lithium content.  The
above pattern is qualitatively  well-reproduced by our models, as well
as by the independent models of \cite{xd09}.

The lithium abundance appears to be  a tight function of the mass, for
stars more  massive than the Sun,  with a few  notable exceptions that
deserve a  closer look.  In  particular, based on  three over-depleted
stars, we suggest  that, at about 1.15 \msun, the  presence of a small
companion may trigger  a large amount of extra  mixing, or perhaps the
initial rotation velocity was higher in these stars. More observations
are needed before drawing firm conclusions about these outliers.

Our  models  qualitatively reproduce  many  observed  features of  the
lithium abundance  as a function of  mass.  However, we  are still far
from achieving  a close match  between observations and models  with a
unique combination of the parameters.

Two stars  in our compilation have chromospheric  activity levels that
are unusually high.  They are  both highly lithium depleted but, apart
from this  circumstance, no strong  relationship between chromospheric
activity and lithium  abundance seems to be present.   Only for one of
the  two  chromospherically  active  stars has  a  projected  rotation
velocity been  measured, which is $\sim  4$ \kms, the  only value that
exceeds  2 \kms\  among the  nine  measurements in  our sample.   This
suggests that  rotation is at the  base of both  lithium depletion and
higher  chromospheric activity  in  this star,  but  that other  extra
mixing processes must also be efficient in some slow rotators.

\begin{acknowledgements}

Valuable comments made by the referee helped to improve the quality of
this paper. This research made use of the SIMBAD database, operated at
CDS,  Strasbourg,  France, and  of  WEBDA,  an  open cluster  database
developed and  maintained by Jean-Claude Mermilliod.   The authors are
grateful  for   the  support  from   FCT/CAPES  cooperation  agreement
n$^{o}$237/09.   G.P.  is supported  by grant  SFRH/BPD/39254/2007 and
and by the  project PTDC/CTE-AST/098528/2008, funded by Funda\c{c}\~ao
para a Ci\^encia e  a Tecnologia (FCT), Portugal.  Research activities
of the Stellar Board at the  Federal University of Rio Grande do Norte
are  supported by  continuous grants  from CNPq  and  FAPERN Brazilian
Agencies.  J.M.   would like to  acknowledge support from  USP, FAPESP
({\em  2010/17510-3}).  J.D.N.   and J.M.   would like  to acknowledge
support from CNPq ({\em Bolsa de Produtividade}).

\end{acknowledgements}

\bibliographystyle{aa}

{}

\Online

\clearpage
\onecolumn

\longtab{1}{

\begin{longtable}{ c c c c c c c c c c c c c }
\caption[M67 data compilation used]{M67 data. See Sects. 2 and 3 for explanations.} 
\\

\hline 

star&V&B-V&pmb&v$_{rad}$& ref & A$_{\rm  orig}$&T$_{\rm  orig}$&A$_{\rm iso-Castro}$&T$_{\rm iso-Castro}$&A$_{\rm  calib}$&T$_{\rm  calib}$&$\Delta$ A\\
&(mag)&(mag)&$\%$& (\kms) &   &  (dex)        & (K)           &  (dex)        & (K)           &  (dex)        & (K)           &dex\\
\hline

\endfirsthead

\multicolumn{13}{c}%
{{\bfseries \tablename\ \thetable{} -- continued from previous page}} \\
\hline 

star&V&B-V&pmb&v$_{rad}$& ref & A$_{\rm  orig}$&T$_{\rm  orig}$&A$_{\rm iso-Castro}$&T$_{\rm iso-Castro}$&A$_{\rm  calib}$&T$_{\rm  calib}$&$\Delta$ A\\

&(mag)&(mag)&$\%$& (\kms) &   &  (dex)        & (K)           &  (dex)        & (K)           &  (dex)        & (K)           &dex\\
\hline

\endhead

\hline 
\multicolumn{13}{|r|}{{Continued on next page}} \\ \hline
\endfoot

\hline 

\multicolumn{13}{l}{References:                                                                                 }\\
\multicolumn{13}{l}{(1)  Canto Martins et al., 2011, A\&A 527, A94;                                             }\\
\multicolumn{13}{l}{(2)  Castro  et al.,  2011, A\&A 526, 17;                                                   }\\              
\multicolumn{13}{l}{(3)  Pasquini et al., 2008, A\&A  489, 677;                                                 }\\
\multicolumn{13}{l}{(4)  Randich et al. 2007, A\&A 469, 163;                                                    }\\                                    
\multicolumn{13}{l}{(5)  Jones, Fischer and Soderblom 1999, ApJ 117, 330, original data;                        }\\ 
\multicolumn{13}{l}{(6)  Jones, Fischer and Soderblom 1999, ApJ 117, 330, correction of literature measurements;}\\     
\multicolumn{13}{l}{(7)  Balachandran 1995, Apj 446, 203.                                                       }\\

\endlastfoot

  Sand. 774    &  12.93   & 0.85   & -- & 33.5    & 1  &  0.0  &5240 &  0.16 &5176 &    0.13   &5253  &   up.lim. \\ 
  YBP   1632   &  12.734  & 0.795  & 100& 33.4    & 1  &  0.0  &5461 &  0.17 &5260 &    0.12   &5398  &   0.2 \\ 
  Sand. 978    &  9.72    & 1.37   & -- & 34.7    & 1  & -1.0  &4260 & -0.73 &4325 &   -0.71   &4283  &   0.3 \\ 
  YBP   1087   &  10.413  & 1.139  & 98 & 33.9    & 1  &  0.0  &4748 &  0.06 &4518 &    0.12   &4623  &   up.lim. \\ 
  Sand. 1016   &  10.3    & 1.26   & -- & 34.6    & 1  & -0.5  &4430 & -0.16 &4487 &   -0.2    &4432  &   0.2     \\ 
  YBP   1248   &  12.639  & 0.615  & 99 & 34.6    & 1  &  1.3  &6020 &  1.38 &6109 &    1.26   &5937  &   0.1     \\ 
  YBP   1479   &  10.462  & 1.127  & 97 & 34.2    & 1  & -0.4  &4750 & -0.14 &4528 &   -0.15   &4644  &   0.2     \\ 
  YBP   829    &  12.891  & 0.935  & 100& 32.9    & 1  &  0.1  &5130 &  0.24 &5128 &    0.25   &5029  &   up.lim. \\ 
  YBP   923    &  12.753  & 0.744  & 100& 32.719  & 1  &  0.8  &5644 &  0.57 &5284 &    0.82   &5541  &   up.lim. \\ 
  YBP   942    &  12.678  & 0.723  & 99 & 37.972  & 1  &  2.7  &5810 &  2.36 &5395 &    2.54   &5601  &   0.1     \\ 
  YBP   973    &  12.944  & 0.904  & 99 & 33.009  & 1  &  0.0  &5170 &  0.14 &5176 &    0.15   &5121  &   0.2     \\ 
  YBP   1060   &  11.465  & 1.04   & 97 & 32.9    & 1  & -0.4  &4820 & -0.19 &4742 &   -0.2    &4801  &   up.lim. \\ 
  YBP   1258   &  12.616  & 0.587  & 99 & 32.7    & 1  &  0.9  &5996 &  1.04 &6081 &    1.03   &6031  &   0.1     \\ 
  YBP   1318   &  12.238  & 0.572  & 98 & 34.641  & 1  &  1.9  &6159 &  1.74 &5929 &    1.85   &6082  &   0.1     \\ 
  YBP   1320   &  12.579  & 0.606  & 99 & 34.322  & 1  &  2.2  &6050 &  2.2  &6039 &    2.15   &5967  &   0.1     \\ 
  YBP   1327   &  11.598  & 1.057  & 98 & 34.6    & 1  & -0.4  &4820 & -0.21 &4775 &   -0.18   &4769  &   up.lim. \\ 
  YBP   1362   &  10.492  & 1.122  & 97 & 33.65   & 1  & -0.5  &4779 & -0.19 &4539 &   -0.22   &4652  &   0.2     \\ 
  YBP   1546   &  12.245  & 0.986  & 96 & 33.978  & 1  & -0.2  &4940 & -0.04 &4909 &   -0.04   &4910  &   up.lim. \\ 
  YBP   1777   &  12.87   & 0.932  & 99 & 34.4    & 1  &  0.4  &5104 &  0.55 &5128 &    0.52   &5037  &   up.lim. \\ 
  YBP   1844   &  12.764  & 0.736  & 91 & 33.2    & 1  &  1.0  &5654 &  0.74 &5284 &    0.99   &5564  &   up.lim. \\ 
  YBP   846    &  12.838  & 0.824  & 100& 33.1    & 1  &  0.0  &5420 &  0.2  &5176 &    0.16   &5321  &   up.lim. \\ 
  YBP   1876   &  12.577  & 0.641  & 99 & 33.921  & 1  &  1.1  &5940 &  0.96 &5727 &    1.06   &5852  &   0.1 \\ 
  Sand. 1607   &  12.62   & 0.56   & -- & 33.7    & 1  &  1.7  &6127 &  1.69 &6095 &    1.71   &6124  &   0.1 \\ 
  YBP   1070   &  12.631  & 0.62   & 100& 33.637  & 1  &  1.2  &6000 &  1.29 &6095 &    1.16   &5920  &   0.1 \\ 
  YBP   1086   &  12.752  & 0.821  & 100& 33.0    & 1  &  0.7  &5429 &  0.62 &5223 &    0.7    &5329  &   0.1 \\ 
  YBP   285    &  14.461  & 0.704  & 98 & 33.951  & 2  &  0.6  &5836 &  0.63 &5767 &    0.64   &5657  &   0.1 \\ 
  YBP   637    &  14.489  & 0.702  & 99 & 34.241  & 2  &  1.5  &5806 &  1.5  &5754 &    1.43   &5663  &   0.1 \\ 
  YBP   1101   &  14.675  & 0.702  & 99 & 32.927  & 2  &  0.6  &5756 &  0.63 &5649 &    0.64   &5663  &   up.lim. \\ 
  YBP   1194   &  14.614  & 0.667  & 99 & 33.673  & 2  &  1.3  &5766 &  1.29 &5688 &    1.35   &5770  &   up.lim. \\ 
  YBP   1303   &  14.641  & 0.677  & 99 & 33.241  & 2  &  1.2  &5716 &  1.22 &5675 &    1.27   &5739  &   0.1 \\ 
  YBP   1304   &  14.731  & 0.723  & 99 & 33.696  & 2  &  0.8  &5704 &  0.82 &5623 &    0.8    &5601  &   up.lim. \\ 
  YBP   1315   &  14.297  & 0.693  & 99 & 32.434  & 2  &  1.8  &5874 &  1.82 &5861 &    1.69   &5691  &   0.1 \\ 
  YBP   1392   &  14.811  & 0.716  & 97 & 34.171  & 2  &  0.6  &5716 &  0.65 &5584 &    0.65   &5622  &   up.lim. \\ 
  YBP   1787   &  14.547  & 0.667  & 98 & 33.386  & 2  &  1.6  &5768 &  1.61 &5727 &    1.65   &5770  &   0.1 \\ 
  YBP   2018   &  14.565  & 0.672  & 98 & 31.74   & 2  &  1.3  &5693 &  1.37 &5714 &    1.4    &5755  &   up.lim. \\ 
  YBP   266    &  13.601  & 0.611  & 100& 33.392  & 3  &  2.5  &6147 &  2.5  &6151 &    2.36   &5950  &   0.1 \\ 
  YBP   288    &  13.857  & 0.637  & 99 & 32.796  & 3  &  2.3  &6004 &  2.34 &6067 &    2.2    &5865  &   0.1 \\ 
  YBP   349    &  14.301  & 0.677  & 93 & 34.132  & 3  &  1.0  &5952 &  0.94 &5861 &    0.86   &5739  &   up.lim. \\ 
  YBP   350    &  13.624  & 0.602  & 99 & 32.778  & 3  &  2.6  &6024 &  2.68 &6137 &    2.57   &5980  &   0.1 \\ 
  YBP   401    &  13.661  & 0.607  & 97 & 32.917  & 3  &  2.4  &6165 &  2.38 &6137 &    2.26   &5963  &   0.1 \\ 
  YBP   473    &  14.443  & 0.699  & 99 & 35.269  & 3  &  1.0  &5919 &  0.91 &5780 &    0.85   &5672  &   up.lim. \\ 
  YBP   587    &  14.107  & 0.646  & 99 & 33.265  & 3  &  2.3  &6077 &  2.21 &5956 &    2.12   &5836  &   0.1 \\ 
  YBP   613    &  13.254  & 0.653  & 99 & 33.274  & 3  &  2.6  &6202 &  2.53 &6095 &    2.32   &5814  &   0.1 \\ 
  YBP   673    &  14.356  & 0.706  & 99 & 32.902  & 3  &  1.4  &5639 &  1.55 &5834 &    1.41   &5652  &   0.1 \\ 
  YBP   689    &  13.12   & 0.663  & 99 & 33.461  & 3  &  2.3  &6093 &  2.26 &6039 &    2.07   &5783  &   0.1 \\ 
  YBP   750    &  13.576  & 0.639  & 100& 33.709  & 3  &  1.5  &5918 &  1.67 &6151 &    1.45   &5859  &   0.1 \\ 
  YBP   769    &  13.478  & 0.641  & 100& 34.414  & 3  &  2.0  &5984 &  2.12 &6151 &    1.9    &5852  &   0.1 \\ 
  YBP   778    &  13.093  & 0.623  & 99 & 33.815  & 3  &  1.7  &6114 &  1.65 &6039 &    1.55   &5911  &   0.1 \\ 
  YBP   809    &  14.959  & 0.737  & 97 & 32.948  & 3  &  1.0  &5667 &  0.86 &5495 &    0.92   &5561  &   up.lim. \\ 
  YBP   851    &  14.113  & 0.617  & 98 & 34.204  & 3  &  1.3  &5948 &  1.31 &5956 &    1.29   &5930  &   0.1 \\ 
  YBP   911    &  14.547  & 0.673  & 99 & 32.534  & 3  &  1.2  &5885 &  1.08 &5727 &    1.1    &5752  &   0.1 \\ 
  YBP   988    &  14.18   & 0.639  & 99 & 32.705  & 3  &  2.0  &5935 &  2.0  &5929 &    1.94   &5859  &   0.1 \\ 
  YBP   1032   &  14.358  & 0.639  & 97 & 34.302  & 3  &  1.6  &5955 &  1.51 &5834 &    1.53   &5859  &   0.1 \\ 
  YBP   1036   &  14.947  & 0.731  & 98 & 33.884  & 3  &  1.4  &5612 &  1.31 &5495 &    1.37   &5578  &   0.1 \\ 
  YBP   1051   &  14.09   & 0.636  & 99 & 32.029  & 3  &  2.2  &6081 &  2.12 &5970 &    2.04   &5868  &   0.1 \\ 
  YBP   1062   &  14.477  & 0.667  & 99 & 33.206  & 3  &  1.7  &5926 &  1.58 &5767 &    1.58   &5770  &   0.1 \\ 
  YBP   1067   &  14.559  & 0.642  & 100& 34.67   & 3  &  1.0  &5929 &  0.87 &5727 &    0.94   &5849  &   0.1 \\ 
  YBP   1075   &  13.712  & 0.674  & 99 & 32.974  & 3  &  0.7  &5871 &  0.94 &6123 &    0.69   &5749  &   0.1 \\ 
  YBP   1088   &  14.492  & 0.659  & 93 & 33.199  & 3  &  1.0  &5890 &  0.91 &5754 &    0.93   &5795  &   up.lim. \\ 
  YBP   1090   &  13.8    & 0.65   & 100& 32.333  & 3  &  2.3  &6086 &  2.31 &6095 &    2.11   &5824  &   0.1 \\ 
  YBP   1129   &  14.171  & 0.624  & 99 & 34.528  & 3  &  2.1  &5959 &  2.09 &5942 &    2.06   &5907  &   0.1 \\ 
  YBP   1137   &  14.873  & 0.698  & 97 & 33.851  & 3  &  1.0  &5741 &  0.87 &5546 &    0.95   &5675  &   up.lim. \\ 
  YBP   1197   &  13.315  & 0.606  & 100& 34.526  & 3  &  2.2  &6207 &  2.14 &6123 &    2.03   &5967  &   0.1 \\ 
  YBP   1247   &  14.144  & 0.609  & 99 & 32.991  & 3  &  2.1  &5994 &  2.06 &5942 &    2.07   &5957  &   0.1 \\ 
  YBP   1334   &  14.403  & 0.68   & 99 & 32.923  & 3  &  2.0  &5957 &  1.89 &5807 &    1.82   &5730  &   0.1 \\ 
  YBP   1387   &  14.098  & 0.626  & 99 & 33.828  & 3  &  2.3  &6090 &  2.22 &5970 &    2.16   &5901  &   0.1 \\ 
  YBP   1458   &  14.977  & 0.739  & 98 & 33.351  & 3  &  1.0  &5640 &  0.87 &5482 &    0.94   &5555  &   up.lim. \\ 
  YBP   1496   &  13.879  & 0.607  & 100& 34.607  & 3  &  2.6  &6173 &  2.52 &6053 &    2.45   &5963  &   0.1 \\ 
  YBP   1504   &  14.171  & 0.625  & 99 & 33.733  & 3  &  2.2  &5934 &  2.21 &5942 &    2.18   &5904  &   0.1 \\ 
  YBP   1514   &  14.777  & 0.721  & 98 & 33.878  & 3  &  1.6  &5613 &  1.59 &5597 &    1.6    &5607  &   0.1 \\ 
  YBP   1587   &  14.163  & 0.641  & 99 & 32.336  & 3  &  2.0  &5975 &  1.98 &5942 &    1.91   &5852  &   0.1 \\ 
  YBP   1622   &  14.156  & 0.632  & 99 & 33.407  & 3  &  2.2  &6043 &  2.13 &5942 &    2.08   &5881  &   0.1 \\ 
  YBP   1716   &  13.299  & 0.619  & 96 & 33.948  & 3  &  2.3  &6030 &  2.36 &6123 &    2.22   &5924  &   0.1 \\ 
  YBP   1722   &  14.13   & 0.601  & 96 & 34.007  & 3  &  2.2  &6007 &  2.16 &5956 &    2.18   &5983  &   0.1 \\ 
  YBP   1735   &  14.332  & 0.661  & 97 & 33.104  & 3  &  0.9  &5959 &  0.83 &5847 &    0.81   &5789  &   0.1 \\ 
  YBP   1758   &  13.207  & 0.653  & 98 & 34.398  & 3  &  2.4  &6221 &  2.29 &6067 &    2.11   &5814  &   0.1 \\ 
  YBP   1768   &  14.404  & 0.656  & 99 & 34.222  & 3  &  2.1  &5844 &  2.07 &5807 &    2.07   &5805  &   0.1 \\ 
  YBP   1788   &  14.441  & 0.663  & 84 & 33.763  & 3  &  1.8  &5886 &  1.72 &5780 &    1.72   &5783  &   0.1 \\ 
  YBP   1852   &  13.962  & 0.613  & 99 & 32.324  & 3  &  2.1  &6009 &  2.11 &6025 &    2.05   &5943  &   0.1 \\ 
  YBP   1903   &  14.733  & 0.689  & 99 & 32.409  & 3  &  1.0  &5609 &  1.01 &5623 &    1.07   &5703  &   up.lim. \\ 
  YBP   1948   &  14.015  & 0.612  & 97 & 33.136  & 3  &  2.4  &6164 &  2.29 &5997 &    2.25   &5947  &   0.1 \\ 
  YBP   1955   &  14.212  & 0.63   & 98 & 32.588  & 3  &  2.2  &5961 &  2.17 &5915 &    2.14   &5888  &   0.1 \\ 
  YBP   901    &  13.422  & 0.554  & 100&  --     & 4  &  2.63 &6191 &  2.57 &6137 &    2.58   &6145  &   0.07\\
  YBP   963    &  12.76   & 0.565  & 100&  --     & 4  &  2.1  &6210 &  2.09 &6194 &    2.03   &6106  &   0.08\\
  Sand. 998    &  13.06   & 0.558  & -- &  --     & 4  &  2.64 &6223 &  2.5  &6053 &    2.56   &6131  &   0.06\\
  YBP   871    &  13.152  & 0.582  & 100&  --     & 4  &  1.86 &6156 &  1.79 &6053 &    1.79   &6048  &   0.16\\
  Sand. 1064   &  14.04   & 0.661  & -- & 32.7    & 5  &  2.14 &5845 &  2.28 &5997 &    2.12   &5789  &   0.1 \\ 
  YBP   713    &  14.172  & 0.714  & 99 & 35.443  & 5  &  2.07 &5800 &  2.2  &5929 &    1.97   &5628  &   0.1 \\ 
  YBP   1105   &  13.946  & 0.59   & 99 & 32.8    & 5  &  2.55 &6094 &  2.5  &6039 &    2.49   &6020  &   0.1 \\ 
  YBP   1397   &  14.009  & 0.622  & 97 & 36.808  & 5  &  2.28 &5972 &  2.32 &6011 &    2.25   &5914  &   0.1 \\ 
  YBP   1342   &  14.285  & 0.65   & 99 & 33.684  & 5  &  1.15 &5810 &  1.24 &5874 &    1.2    &5824  &   up.lim. \\ 
  Sand. 1057   &  14.3    & 0.668  & -- & 34.5    & 5  &  1.75 &5818 &  1.82 &5861 &    1.75   &5767  &   0.1 \\ 
  YBP   1486   &  13.864  & 0.574  & 94 &  --     & 6  &  2.63 &6178 &  2.54 &6067 &    2.54   &6075  &   0.1 \\ 
  YBP   890    &  13.179  & 0.59   & 100& 34.212  & 6  &  1.48 &6153 &  1.43 &6067 &    1.4    &6020  &   up.lim. \\ 
  YBP   961    &  13.19   & 0.576  & 100& 34.59   & 6  &  2.37 &6151 &  2.31 &6067 &    2.31   &6068  &   0.1 \\ 
  YBP   1680   &  13.646  & 0.645  & 100& 35.428  & 6  &  2.45 &5934 &  2.6  &6137 &    2.38   &5840  &   0.1 \\ 
  Sand. 1055   &  13.8    & 0.593  & -- &  --     & 6  &  2.43 &6116 &  2.41 &6095 &    2.35   &6010  &   0.1 \\ 
  YBP   1456   &  12.841  & 0.943  & 100&  --     & 7  &  0.5  &5000 &  0.75 &5105 &    0.67   &5008  &   up.lim. \\ 
  YBP   877    &  12.732  & 0.583  & 99 &  --     & 7  &  1.9  &6205 &  1.89 &6194 &    1.79   &6044  &   up.lim. \\ 
  YBP   891    &  11.401  & 1.089  & 98 &  --     & 7  &  0.0  &4740 &  0.2  &4731 &    0.18   &4710  &   up.lim. \\ 
  YBP   1017   &  12.361  & 0.962  & 96 & 33.639  & 7  &  0.5  &4905 &  0.7  &4931 &    0.74   &4962  &   up.lim. \\ 
  YBP   1024   &  9.586   & 1.443  & 98 & 31.655  & 7  & -1.0  &4010 & -1.01 &4295 &   -0.83   &4176  &   up.lim. \\ 
  YBP   1337   &  12.878  & 0.972  & 100&  --     & 7  &  0.9  &4850 &  1.35 &5105 &    1.18   &4940  &   0.1 \\ 

\end{longtable}                                                                           

}

\longtab{2}{

\begin{longtable}{c c c c c c c c c c c c}
\caption[Hyades data compilation used]{Hyades data}
\\

\hline 

   vB  &   V   &   B-V &  \teff   &log g & Mass  &   EW &    $\Delta$EW & A(Li) &  $\Delta$ A & EW  ref & phot ref \\
\hline

\endfirsthead

\multicolumn{12}{c}%
{{\bfseries \tablename\ \thetable{} -- Data in the previous page}} \\
\hline 

%   vB  &   V   &   B-V &  \teff   &log g & Mass  &   EW &   $\Delta$EW & A(Li) &   $\Delta$ A & EW ref & phot ref\\

\endhead

\hline 
\multicolumn{12}{|r|}{{References on next page}} \\ \hline
\endfoot
\hline

\hline 

\multicolumn{12}{l}{References for EW measurements:                        }\\
\\
\multicolumn{12}{l}{(Ra)     Randich et al. 2007, A\&A, 469, 163;          }\\ 
\multicolumn{12}{l}{(T)      Thorburn et al., 1993, ApJ, 415, 150;         }\\
\multicolumn{12}{l}{(S)      Soderblom et al., 1990, AJ, 99, 595;          }\\
\multicolumn{12}{l}{(BB)     Boesgaard \& Budge, 1988, ApJ, 332, 410;      }\\                                          
\multicolumn{12}{l}{(BT)     Boesgaard \& Tripicco, 1986, ApJ, 302, L49;   }\\
\multicolumn{12}{l}{(D)      Duncan \& Jones, 1983, ApJ, 271, 663;         }\\              
\multicolumn{12}{l}{(Re)     Rebolo \& Beckman, 1988, A\&A, 201, 267;      }\\     
\hline
\\
\hline
\multicolumn{12}{l}{References for the photometry:                         }\\
\\
\multicolumn{12}{l}{(Bal.)   Balachandran, 1995, ApJ, 446, 203;            }\\
\multicolumn{12}{l}{(Joner)  Joner et al., 2006, AJ, 132, 111;             }\\

\endlastfoot

\hline
\hline
     2 &  7.78 & 0.617 & 5916 & 4.36 & 1.140 & 89.0 &  2 & 2.73 &  0.03 &  T & Bal.  \\  
     3 &  8.88 & 0.848 & 5099 & 4.48 & 0.847 &$\leq$25.0&&$\leq$ 1.37 & &  D & Bal.  \\  
     4 &  5.97 & 0.341 & 6897 & 3.99 & 1.524 &$\leq$3.0 &&$\leq$ 1.79 & &  T & Bal.  \\  
     6 &  5.95 & 0.348 & 6861 & 3.96 & 1.513 & 54.8 &  3 & 3.10 &  0.04 & BB & Joner \\  
     8 &  6.37 & 0.419 & 6544 & 4.02 & 1.411 &$\leq$4.0 &&$\leq$ 1.69 & & BB & Bal.  \\  
    10 &  7.85 & 0.589 & 6000 & 4.42 & 1.182 & 82.0 &  2 & 2.74 &  0.03 &  T & Bal.  \\  
    13 &  6.62 & 0.420 & 6542 & 4.12 & 1.410 &$\leq$5.0 &&$\leq$ 1.79 & & BT & Bal.  \\  
    14 &  5.73 & 0.355 & 6825 & 3.87 & 1.502 & 65.0 &  3 & 3.18 &  0.04 & BT & Bal.  \\  
    15 &  8.09 & 0.658 & 5703 & 4.39 & 1.037 & 57.0 &  2 & 2.32 &  0.03 &  T & Bal.  \\  
    17 &  8.46 & 0.696 & 5539 & 4.47 & 0.971 & 42.0 &  2 & 2.03 &  0.05 &  T & Bal.  \\  
    18 &  8.06 & 0.638 & 5779 & 4.41 & 1.072 & 74.0 &  2 & 2.52 &  0.04 &  T & Bal.  \\  
    19 &  7.14 & 0.512 & 6231 & 4.22 & 1.290 & 82.4 &  3 & 2.91 &  0.03 & BB & Bal.  \\  
    20 &  6.32 & 0.399 & 6624 & 4.03 & 1.438 & 87.6 &  3 & 3.21 &  0.03 & BB & Bal.  \\  
    21 &  9.15 & 0.816 & 5207 & 4.63 & 0.872 &$\leq$3.0 &&$\leq$ 0.58 & & Ra & Bal.  \\  
    26 &  8.63 & 0.743 & 5437 & 4.51 & 0.936 & 13.0 &  2 & 1.39 &  0.07 &  T & Bal.  \\  
    27 &  8.46 & 0.715 & 5506 & 4.46 & 0.960 & 23.0 &  2 & 1.72 &  0.04 &  T & Bal.  \\  
    31 &  7.47 & 0.566 & 6067 & 4.29 & 1.215 & 95.0 &  2 & 2.88 &  0.02 &  T & Bal.  \\  
    36 &  6.80 & 0.441 & 6463 & 4.17 & 1.382 &  5.9 &  3 & 1.81 &  0.18 & BT & Bal.  \\  
    37 &  6.61 & 0.405 & 6600 & 4.14 & 1.430 & 10.3 &  3 & 2.14 &  0.12 & BT & Bal.  \\  
    38 &  5.72 & 0.320 & 7019 & 3.93 & 1.560 & 19.6 &  3 & 2.70 &  0.14 & BT & Bal.  \\  
    42 &  8.86 & 0.759 & 5414 & 4.59 & 0.929 & 11.0 &  2 & 1.30 &  0.08 &  T & Bal.  \\  
    44 &  7.19 & 0.450 & 6433 & 4.31 & 1.371 & 20.4 &  3 & 2.34 &  0.07 & BT & Bal.  \\  
    46 &  9.11 & 0.867 & 5028 & 4.55 & 0.832 &$\leq$4.0 &&$\leq$ 0.55 & &  T & Bal.  \\  
    48 &  7.14 & 0.521 & 6203 & 4.21 & 1.278 & 91.0 &  2 & 2.95 &  0.02 &  T & Bal.  \\  
    51 &  6.97 & 0.443 & 6457 & 4.23 & 1.380 &  6.5 &  3 & 1.84 &  0.17 & BT & Bal.  \\  
    49 &  8.24 & 0.585 & 6012 & 4.58 & 1.188 & 57.0 &  2 & 2.55 &  0.03 &  T & Bal.  \\  
    59 &  7.49 & 0.543 & 6136 & 4.33 & 1.248 & 82.0 &  2 & 2.84 &  0.03 &  T & Bal.  \\  
    61 &  7.38 & 0.514 & 6224 & 4.32 & 1.287 &110.8 &  3 & 3.08 &  0.03 & BB & Bal.  \\  
    64 &  8.12 & 0.657 & 5707 & 4.40 & 1.039 & 59.0 &  2 & 2.34 &  0.03 &  T & Bal.  \\  
    65 &  7.42 & 0.535 & 6160 & 4.31 & 1.259 &106.0 &  2 & 3.00 &  0.03 &  T & Bal.  \\  
    66 &  7.51 & 0.555 & 6100 & 4.32 & 1.231 & 73.0 &  2 & 2.75 &  0.03 &  T & Bal.  \\  
    73 &  7.84 & 0.609 & 5940 & 4.39 & 1.152 & 85.0 &  2 & 2.72 &  0.03 &  T & Bal.  \\  
    77 &  7.05 & 0.500 & 6263 & 4.20 & 1.304 & 19.0 &  6 & 2.20 &  0.13 & Re & Bal.  \\  
    76 &  9.20 & 0.759 & 5392 & 4.72 & 0.922 & 16.0 &  2 & 1.45 &  0.06 &  T & Bal.  \\  
    78 &  6.92 & 0.453 & 6422 & 4.20 & 1.367 & 32.6 &  3 & 2.56 &  0.05 & BB & Bal.  \\  
    79 &  8.93 & 0.827 & 5171 & 4.53 & 0.863 &$\leq$3.0 &&$\leq$ 0.56 & & T & Joner  \\  
    81 &  7.10 & 0.470 & 6365 & 4.25 & 1.345 & 15.6 &  3 & 2.18 &  0.08 & BT & Bal.  \\  
    85 &  6.51 & 0.426 & 6519 & 4.07 & 1.402 &$\leq$6.0 &&$\leq$ 1.85 & & BB & Bal.  \\  
    86 &  7.05 & 0.463 & 6388 & 4.24 & 1.354 & 20.7 &  3 & 2.32 &  0.07 & BT & Bal.  \\  
    87 &  8.58 & 0.743 & 5437 & 4.49 & 0.936 & 12.0 &  2 & 1.36 &  0.07 &  T & Bal.  \\  
    88 &  7.75 & 0.554 & 6103 & 4.42 & 1.232 & 89.0 & 10 & 2.86 &  0.06 &  S & Joner \\  
    90 &  6.40 & 0.413 & 6568 & 4.05 & 1.419 &$\leq$3.0 &&$\leq$ 1.58 & & BB & Bal.  \\  
    92 &  8.66 & 0.741 & 5443 & 4.52 & 0.938 & 15.0 &  2 & 1.46 &  0.06 &  T & Bal.  \\  
    93 &  9.40 & 0.883 & 4968 & 4.64 & 0.820 &$\leq$3.0 &&$\leq$ 0.29 & &  T & Bal.  \\  
    94 &  6.62 & 0.431 & 6499 & 4.11 & 1.395 &$\leq$2.2 &&$\leq$ 1.40 & & BT & Bal.  \\  
    97 &  7.93 & 0.634 & 5793 & 4.36 & 1.079 & 84.0 &  2 & 2.60 &  0.05 &  T & Bal.  \\  
    99 &  9.38 & 0.851 & 5090 & 4.68 & 0.845 &  4.0 &  2 & 0.59 &  0.13 &  T & Bal.  \\  
   101 &  6.65 & 0.433 & 6493 & 4.12 & 1.393 &$\leq$3.0 &&$\leq$ 1.53 & & BB & Bal.  \\  
   105 &  7.53 & 0.575 & 6042 & 4.31 & 1.203 & 87.0 &  2 & 2.81 &  0.02 &  T & Bal.  \\  
   109 &  9.40 & 0.817 & 5203 & 4.73 & 0.871 &  5.0 &  1 & 0.74 &  0.09 & Ra & Bal.  \\  
   116 &  9.01 & 0.821 & 5191 & 4.57 & 0.868 &$\leq$5.0 &&$\leq$ 0.73 & &  T & Bal.  \\  
   118 &  7.74 & 0.580 & 6026 & 4.39 & 1.195 & 77.0 &  2 & 2.72 &  0.03 &  T & Bal.  \\  
   127 &  8.92 & 0.710 & 5520 & 4.65 & 0.964 & 18.0 &  2 & 1.62 &  0.05 &  T & SIMBAD\\  
   153 &  8.91 & 0.859 & 5062 & 4.48 & 0.839 &  8.0 &  2 & 0.82 &  0.10 &  T & Bal.  \\  
   121 &  7.29 & 0.504 & 6256 & 4.29 & 1.301 &114.9 &  3 & 3.12 &  0.02 & BT & Bal.  \\  
   124 &  6.25 & 0.501 & 6261 & 3.88 & 1.303 &  9.0 &  3 & 1.86 &  0.13 & BT & Joner \\  
   128 &  6.75 & 0.450 & 6433 & 4.14 & 1.371 & 14.0 &  3 & 2.17 &  0.09 & BT & Bal.  \\  
   180 &  9.10 & 0.853 & 5081 & 4.57 & 0.843 &  5.0 &  2 & 0.65 &  0.13 &  T & Bal.  \\  
   187 &  8.60 & 0.776 & 5352 & 4.46 & 0.912 & 17.0 &  1 & 1.44 &  0.04 & Ra & Joner \\ 
\end{longtable}
}

\end{document}